# Optical probing of the metal-to-insulator transition in a 2D high mobility electron gas

F Dionigi[1#], F Rossella[1], V Bellani[1*], M Amado[2,3], E Diez[3], K Kowalik[4§],

G Biasiol[5] and L Sorba[6]

[1]*Dipartimento di Fisica "A. Volta" and CNISM, Università degli Studi di Pavia, 27100 Pavia, Italy*
[2]*GISC and Departamento de Física de Materiales, Universidad Complutense, 28040 Madrid, Spain*
[3]*Laboratorio de Bajas Temperaturas, Universidad de Salamanca, 37008 Salamanca, Spain*
[4]*Laboratoire National des Champs Magnétiques Intenses, CNRS, 38042 Grenoble, France*
[5]*Istituto Officina dei Materiali CNR, Laboratorio TASC, 34149 Trieste, Italy*
[6]*NEST, Istituto Nanoscienze-CNR and Scuola Normale Superiore, 56126 Pisa, Italy*
Present addresses: [#] *Institut for Fysik, Danmarks Tekniske Universitet, 2800 Lyngby, Denmark.*
[§] *Department für Physik, Ludwig Maximilians Universität, 80539 München Germany.*

**Abstract:** We study the quantum Hall liquid and the metal-insulator transition in a high mobility two dimensional electron gas, by means of photoluminescence and magneto-transport. In the integer and fractional regime at $\nu > 1/3$, analyzing the emission energy dispersion we probe the magneto-Coulomb screening and the hidden symmetry of the electron liquid. In the fractional regime above above $\nu = 1/3$ the system undergoes the metal-to-insulator transition, and in the insulating phase the dispersion becomes linear with evidence of an increased renormalized mass.

## 1. Introduction

The two dimensional electron gas (2DEG) in the quantum Hall regime is the subject of intense experimental and theoretical studies since this system is still revealing new surprising properties. Indeed recent studies evidence unexpected structures in the density of state spectra of the 2DEG in the fractional regime which suggest that there exist core features of the system that have yet to be understood [1]. Other experimental works [2] are also unveiling the properties of quantum Hall states with non-Abelian statistic which has been proposed as the basis for a fault-tolerant topological quantum computer. Recent studies showed that in 2DEG with intermediate electron density of about $10^{11}$ cm$^{-2}$ and mobility between $1 \times 10^6$ cm$^2$/Vs and $3 \times 10^6$ cm$^2$/Vs the interplay between disorder and Coulomb interactions can reveal interesting interaction-related effects like energetic anomalies of the charged excitons [3] and peculiar fractional states like the $\nu = 3/8$ [4].

The objective of this work is to probe the 2DEG at intermediate density by means of the complementary use of photoluminescence and transport measurements, in order to unveil new properties of the metal-to-insulator transition for the quantum Hall system in the fractional regime, focusing on the insulating phase where we find evidence of an increased renormalized mass. Indeed the use of complementary PL and transport experiments proves to be an effective tool for the investigation of the many-body interactions of the 2DEG electrons [5,6].

In an asymmetric modulation doped QW the photo-excited holes are able to probe the plasma of interacting electrons, since the holes weakly affect the 2DEG when the separation between electrons and holes is large. In particular a magnetic field $B$ normal to the QW enhances many-body effects and gives access to quantum phenomena such as the integer and fractional quantum Hall effect (IQHE and FQHE, respectively). The photoluminescence (PL) of a modulation doped QW in magnetic field strongly depends on the sheet electron density $n$ of the 2DEG. Three regimes are usually observed: low density ($n \leq 10^{10}$ cm$^{-2}$), where the PL emission is mainly due to excitonic recombination; intermediate density or also called dilute regime ($n \sim 10^{11}$ cm$^{-2}$), where charged excitons, also named trions, dominate the spectrum; and high density ($n > 10^{11}$ cm$^{-2}$) where the recombination of the photo-excited holes is strongly influenced by the whole 2DEG [7].

When the magnetic field exceeds a critical value $B_c$, the system from a quantum Hall liquid exhibiting the quantum Hall plateau becomes insulating, characterized by a diverging longitudinal resistance, as the temperature vanished. This transition to the insulating state, called also two-dimensional metal-insulator transition, has been studied by electrical transport, and has been found that for low mobility samples the transition occurs beyond the $v = 1$ quantum Hall state (being $v = nh/(eB)$ the filling factor) while in high mobility GaAs-AlGaAs samples the transition takes place beyond the $v = 1/3$ state or the $v = 1/5$ state [8]. This liquid-to-insulator transition has been successively studied in details by electrical transport, in low mobility samples like InGaAs/InP and Ge/SiGe and in GaAs-AlGaAs high mobility structures [8] and recently in graphene [9], while the high field regime has been studied optically by PL concerning the emission energy and line-shape [6,10]. The nature of the insulating state is still an open question, since the Anderson localization (due to disorder) and the Mott localization (driven by the strong correlation between the interacting electrons) are the two candidates for the explanation of the phenomenon [11,12]. In a recent picture proposed by Camjayi *et al*. [11], the lattice sites represent the precursors, in the fluid phase, of vacancies and interstitials in the Wigner crystal phase, and the key aspects of the phase diagram resulting within this picture reconcile the early view points of Wigner and Mott. In this scenario the metal-insulator transition in the 2DEG could be attributed to strong correlations that are effectively enhanced even far away from integer filling, due to incipient charge ordering driven by non-local Coulomb interactions.

The intermediate value of $n$ and the moderate-high mobility $\mu$ of our 2DEG allow: (i) to observe a deviation from the single particle model with Landau level linear dispersion at low field, and the behavior typical of trions as well as of the sea of strong interacting electrons in the intermediate field regime; (ii) to probe the metal-to-insulator transition in the 2DEG at high magnetic field. The quantitative analysis of the experimental data gives the value of the renormalized mass in the insulating phase at critical density. The results indicate that this phase is compatible with the incipient charge ordering driven by non-local Coulomb interactions presented in recent theoretical models [11,12].

## 2. Experimental and results

The experiments were performed on a 20 nm-thick single-sided modulation-doped GaAs/AlGaAs single QW with carrier density $n = 1.8 \times 10^{11}$ cm$^{-2}$ and mobility $\mu = 1.6 \times 10^{6}$ cm$^2$ V$^{-1}$ s$^{-1}$. Indium contacts were placed with a Hall bar geometry on a rectangular sample of milli-metric size. The sample was excited with the light from a Ti:sapphire laser at the energy of 1.748 eV, with an excitation power smaller than 300 $\mu$W and a laser spot size of ~ 1 mm$^2$.

At zero magnetic field the PL of the sample consists in a broad peak whose width is approximately equal to the Fermi energy, which is typical for a 2DEG with density above ~ $10^{11}$ cm$^{-2}$ [13]. Applying a magnetic field additional emission lines come out from the high energy side of this peak, due to the recombination of electrons and photo-created holes from the quantized Landau levels (LLs).

In Fig. 1 we report the contour plot of the emission at $T$ ~ 40 mK at low field. The strongest emission line, labeled (0,0) in the notation of Rashba and coworkers [13], derives from the recombination of the 2DEG electrons and photo-generated holes in their fundamental LLs in conduction and valence band, respectively. We can see emissions from higher LLs on the high energy side of (0,0), which are marked with the corresponding index of electrons and holes LLs. We notice that the emissions labeled as (2,2) from the third LLs and (3,3) from the fourth ones follow very well the single-particle linear dispersion. Moreover we can see shifts in the position of the (0,0) emission at magnetic fields close and slightly smaller than the ones corresponding to integer values of the $v$, related the discontinuities of Fermi level. We also observe that at the field of ~ 3.8 T the transition from (1,0) disappears; this is due to the emptying of the electron LL which takes place for $v > 2$ since the Fermi levels drops to the first LL, and allows us identify the position of $v = 2$ at the field $B$ ~ 3.8 T that we label $B_{v=2}$. On the low energy side of the (0,0) line we identify the emission deriving from a shake-up (SU) process that also follows a linear dispersion. This shake-up process is explained as a single particle excitation in which an electron is excited from a LL to an higher LL with the simultaneous absorption of energy from another electron of the 2DEG which is recombining with a photo-created hole. This loss of energy, namely $(N+1)\hbar\omega_{c\text{-}e}$ (being $\omega_{c\text{-}e}$ the electronic cyclotron frequency and $N$ the LL index ) is the cause of the red shift of the energy of the emitted photon. Other authors report that at higher electron density the many-body interactions transform this excitation into a magneto-plasmon, that has a more complicated dispersion and a greater energy shift ~1.6 $\hbar\omega_c$ [14]. In our measurements the red-shift of the shake-up is reduced respect to the expected one of the same amount of the reduction of the blue shift of the (1,0), indicating that these two transitions involve, respectively, the loss and gain of the same energy. Even if, theoretically, only transitions that satisfy the selection rule $\Delta N = 0$ are allowed, small coupling can lead to the presence of several forbidden transitions in the experimental spectra [15].

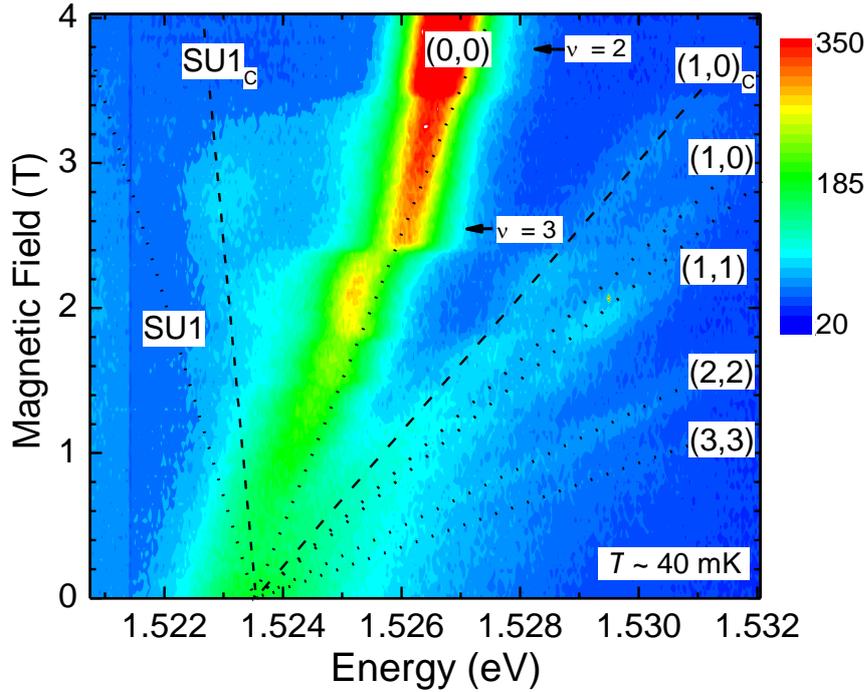

**Fig. 1.** Contour plot of the emission of the 2DEG at $T \sim 40$ mK. The dotted lines are the calculated dispersions using Landau level behavior, while the dashed lines include a corrective coefficient for the SU and (1,0).

As previously mentioned, for low fields the PL emission approximately follows a linear dispersion close to the not interacting single particle energy dispersion given by the formula of the energy of the LLs, $E_L = \hbar (N + 1/2)(\omega_e + \omega_h)$ where $N = 0,1,2,3...$ is an integer that labels the LLs and $\omega_{e,h}$ are the cyclotron frequencies defined as $eB/m_{e,h}^*$ with $m_e^*$ and $m_h^*$ the effective mass for electrons and holes, respectively. The dotted lines in Fig. 1 indicate the theoretical position of the main transitions. We can see that the theoretical dispersion for the $(N,N)$ with $N = 1,2$ and 3 correspond well to observed transitions, while in the case of the SU process and of the (1,0) these theoretical dispersion over-estimate the experimental ones. We found that if these dispersions for the SU and (1,0) are corrected by the same multiplicative factor ~ 0.7, the resulting dispersions (dashed lines of Fig. 1), labeled $SU_C$ and $(1,0)_C$ well describe the emissions closer to the (0,0), at his left and right energy side, respectively. The fact that the corrective factor is the same for the SU and (1,0), i.e. that SU and (1,0) lie at the same energetic distance with respect to the (0,0), is coherent with the fact that SU and (1,0) are energetically related to the (0,0) by term of the order of $\omega_e$.

In Fig. 2 we report the energy position of the emission (0,0) as a function of $B$. On the whole, we can distinguish two different regions, a low field region for $B < B_{v=2}$ where the (0,0) emission energy approximately follows the LL dispersion, and a high field region for $B > B_{v=2}$ where the dispersion deviated and stays well below the linear behavior $1/2(\hbar\omega_{c-e} + \hbar\omega_{c-h})$. At low field the experimental dispersion can be described introducing the magneto-Coulomb energy $E_C = e^2/(\varepsilon l_0)$ [13] that takes into account the Coulomb interaction and corrects the energy dispersion of the (0,0), SU, (1,0) and (1,1). In this expression $\varepsilon$ is the permittivity of GaAs and $l_0 = (\hbar/eB)^{1/2}$ is the magnetic length, and fit of the energy dispersion curve at low field is improved multiplying $E_C$ by a suitable numeric coefficient [13].

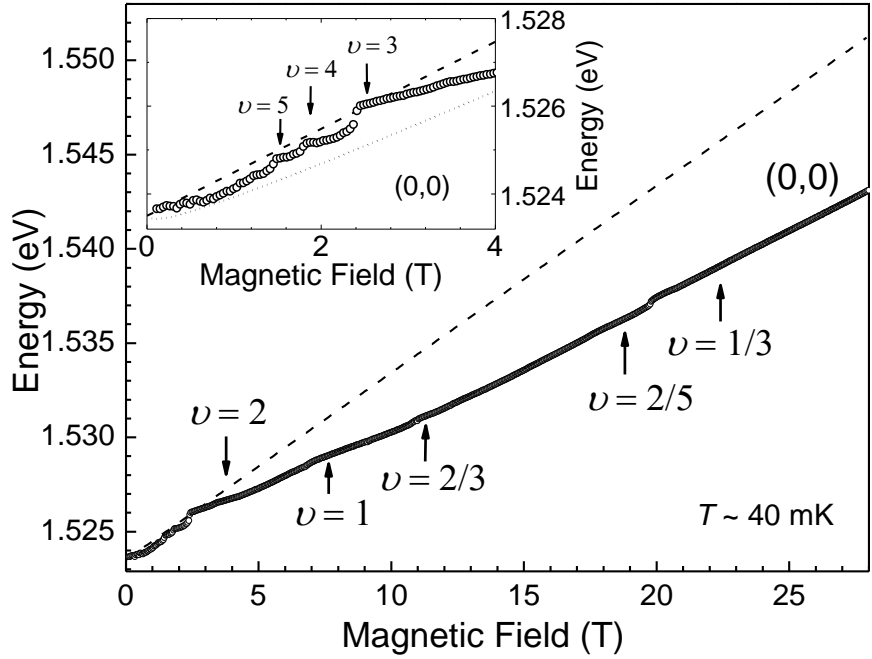

**Fig. 2.** Energy position of the (0,0) PL line versus $B$ at $T \sim 40$ mK. The dashed-line is the Landau behavior. The low field region is expanded in the inset, where the dotted line takes into account the $E_c$ contribution.

At field higher than $B_{v=2}$ the energy dispersion of (0,0) emission strongly deviates from the theoretical (0,0) linear emission, with two small inflections/shifts at $v = 2/3$ and between $v = 2/5$ and $v = 1/3$, which have been also observed in other works [5] related to the FQHE and interpreted as originated by the interaction of the two flux quanta in a composite fermion with a recombining exciton. These features are typically observed at high electron density and disappear decreasing the density as shown by Bar-Joseph *et al.* [14]. In our sample we observed and studied the suppression of the PL emission due to interacting composite fermions in fractional quantum Hall regime [4], where the shifts and suppressions of PL imply a significant correlations and a many body interaction between the electrons of the 2DEG. Actually the magnetic field range between $B_{v=2/5}$ and $B_{v=1/3}$ is interesting not only because of the shifts in energy, but also because we observe another change in the overall behavior of the energy dispersion, as it will be discussed in the following.

In Fig. 3 (a) we report the difference between the experimental energy dispersion and its quadratic best-fit calculated from 3.5 T till to the maximum field of 28 T. We can see that apart the features at $v = 1$, 2/5 and 1/3, the curve lays around zero indicating that the quadratic behavior is good up to 19 T. A quadratic best-fit in the reduced region (3.5 T $< B <$ 19 T) gives indeed a better agreement. Actually, the quadratic dispersion is typical for electron-hole bound states like neutral or charged excitons (trions). Indeed, Yoon *et al.* [7] observed a similar behavior with a dispersion close to the one of the singlet state of the trion, either at low and intermediate density. A transition from linear to quadratic dispersion at $B_{\nu=2}$ has been observed by several authors with samples in the intermediate regime of density and symmetric QW, and the common explanation invokes the occurrence of the so called hidden symmetry, which requires the conditions of strictly 2D system in the limit of infinite $B$, occupation of the lowest LL and charge symmetric interaction of electrons and holes [11]. This hidden symmetry entails that the only possible transition is the excitonic recombination at $k = 0$ at the energy that the exciton would have in absence of the 2DEG liquid. Usually asymmetric QW do not exhibit a clear kink at $\nu = 2$ (it has been argued that the hidden symmetry can't hold in this case) but this is not in contrast with our finding, since as demonstrated by Hayne *et al.* [16], also for an asymmetric QW it's possible to observe this strong shift as well as the quadratic behavior if the band has been flattened by photo-irradiation or when the width of the QW is not too wide.

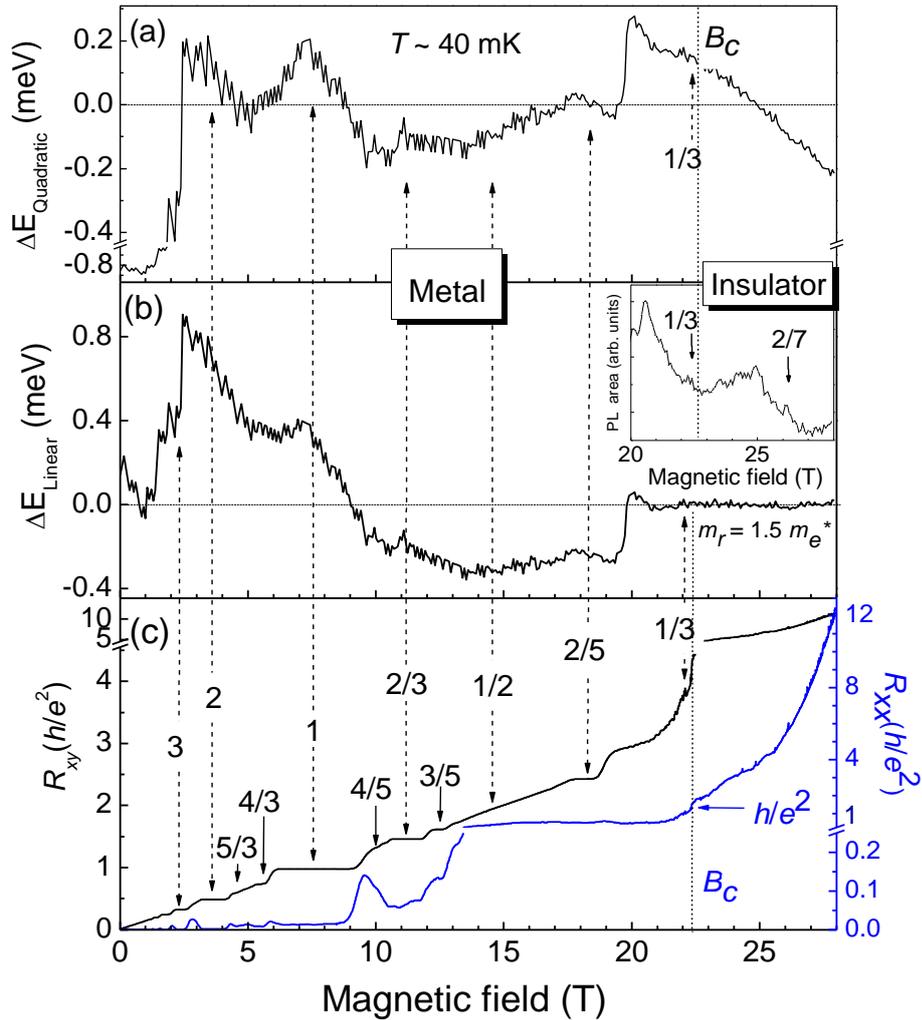

**Fig. 3.** The upper (middle) panel reports the energy $\Delta E_{Quadratic}$ ($\Delta E_{Linear}$) obtained subtracting from the experimental dispersion its quadratic (linear) best-fit calculated above 19 T (from 3.5 T till of 28 T); The inset of the middle panel shows the PL area above $\nu = 1/3$. The bottom panel shows $R_{xx}$ and $R_{xy}$ versus B. The measurements are taken at $T \sim 40$ mK.

Recently Blokland *et al.* [10] studied the photoluminescence emission in the fractional quantum Hall regime, analyzing the fine structure arising from the recombination of fractional charged elementary excitations on the 2DEG and the itinerant valence band holes. The experiments reveals a splitting of the emission band in three lines at $\nu = 2/3$ and in the region $2/5 > \nu > 1/3$, with a lowest-energy emission between $\nu = 2/5$ and $\nu = 1/3$ from magneto-roton assisted transitions from the ground state of the photo-excited fractional quantum Hall system. While our 2DEG is placed in a QW, the samples studied by Blokland *et al.* are single heterojunctions (HJ) with electron density similar to that of our sample, but with higher mobility ($\mu > 5 \times 10^6$ cm$^2$ V$^{-1}$ s$^{-1}$). The different potential profile of HJ and QW samples and the different mobility can lead to a different interplay between many body interactions and localization that results in the diverse dispersions and line splitting observed in the experiments.

At $B$ above ~ 21 T the emission follows another dispersion law that we find out to be linear as can be seen in Fig. 3 (b) where we plotted the curve obtained by subtracting from the experimental dispersion its linear best-fit, calculated above 21 T; the slope of the linear fit is 0.71 times the slope of the theoretical LL line reported in Fig. 2 (dashed line), and both lines intercept approximately the same zero field emission energy ~1.524 eV. From the factor 0.71 which accounts for the changed slope of the linear dispersion above $v = 1/3$, we obtain a $m_r$ that is 1.5 times the band electron mass, a value which is in good agreement with temperature dependent conductivity experiments [18]. Usually the exciton exhibits the quadratic behavior from $B = 0$ up to a field $B_{Ryd}$ at which the LL energy is equal to the excitonic Rydberg energy, and in our case $B_{Ryd}$ is calculated to be 16.1 T. However if we calculate the crossing between the two-dimensional effective Rydberg and the LL energy weighted with the factor 0.71, we obtain $B = 22.5$ T, a value very close to that at which we observe the beginning of the linear behavior.

The change in the behavior of the emission energy from the 2DEG can be interpreted also in the light of the results of magneto transport experiments that we have realized on the same sample, measuring the longitudinal ($R_{xx}$) and Hall ($R_{xy}$) resistances. In Fig. 3 (c), $R_{xx}$ and $R_{xy}$ measured at $T$ ~ 40 mK are reported, and several fractional states can be observed, together with the integer ones. At fields above $B_{v=1/3}$ the resistance $R_{xx}$ increases rapidly, showing that the system becomes insulating. Indeed the transition from the quantum Hall liquid to the insulator is interpreted as a quantum phase transition, characterized by a critical magnetic field $B_c$ [8] at which $R_{xx} = h/e^2$. In the insulating phase ideally the $R_{xx}$ shows an exponential growth, while the $R_{xy}$ is expected to remain quantized (in the low mobility samples at the value $h/e^2$ above $v = 1$, and in the high mobility samples at the value of $h/3e^2$ above $v = 1/3$). However, a small Hall-contact results in a growth of $R_{xy}$ which moves away from its quantized value, as it appears in our Fig. 3 (c), due to a contribution of the exponentially growing $R_{xx}$ [8]. This deviation of the $R_{xy}$ from the quantized value affects either the measurements in GaAs-AlGaAs high mobility samples and in low mobility InGaAs-InP and Ge-SiGe, and to a certain degree can be circumvented by symmetrizing the measurement, *i.e.* reversing the magnetic field $B$ orientation as the contribution of $R_{xx}$ is symmetric in $B$ as opposed to antisymmetric for $R_{xy}$ [8]. Moreover we see that $R_{xx}$ does not really vanish in correspondence to the Hall plateaus in $R_{xy}$, due to parallel conduction effects. However, we notice that the optical results are not affected by the parallel conduction, since are obtained from the analysis of the (0,0) emission of the 2DEG in the QW, whereas the carriers involved in the parallel conduction lay in regions different from the QW.

## 3. Conclusions

In conclusion, we studied the optical emission and the transport in a high mobility two dimensional electron gas at intermediate density in the quantum Hall regime. At low field for $v > 2$, the emission follows a Landau dispersion with a screened magneto-Coulomb contribution while for $2 > v$ the hidden symmetry manifests. For the value of mobility and electron density of the sample, the interplay between disorder and Coulomb interactions reveal interesting interaction-related effects. In particular beyond $v = 1/3$ when the systems enters in the insulating phase, we find that the energy dispersion of the emission becomes linear, with the value of the increased renormalized mass in agreement with experiments on strongly interacting electron systems at critical density. These results show that the transition to the insulating phase could be attributed to the incipient charge ordering driven by non-local Coulomb interactions revealed by recent theoretical models in which the lattice sites represent the precursors, in the fluid phase, of vacancies and interstitials in the Wigner crystal phase.

**Acknowledgements**

This work has been financially supported by the Cariplo Foundation (project *QUANTDEV*), MEC FIS2009-07880, PPT310000-2009-3, JCYL SA049A10-2 and by EuroMagNET under the EU contract nº 228043.